\documentclass[a4paper,11pt]{article}
\usepackage{jheppub} 
\usepackage{xcolor}
\usepackage{bm}
\usepackage{bbold}
\usepackage{tikz}
\usepackage[compat=1.1.0]{tikz-feynman}
\tikzset{
    cross/.style={path picture={\draw[black]
        (path picture bounding box.south east) -- (path picture bounding box.north west)
        (path picture bounding box.south west) -- (path picture bounding box.north east);}}
}
\usepackage{enumitem}

\arxivnumber{2509.17239} 

\title{\boldmath Renormalization of massless fields in the $(1,0)\oplus(0,1)$ representation. }

\author[a]{Armando de la C. Rangel-Pantoja,}
\emailAdd{adlc.rangelpantoja@ugto.mx}
\author[a]{M. Napsuciale}
\emailAdd{mauro@fisica.ugto.mx}
\author[a,b,c]{Carlos A. Vaquera-Araujo}
\emailAdd{vaquera@fisica.ugto.mx}
\affiliation[a]{Departamento de F\'isica, DCI, Campus Le\'on, Universidad de Guanajuato,\\ 
Loma del Bosque 103, Lomas del Campestre C.P. 37150, Le\'on, Guanajuato, M\'exico}
\affiliation[b]{Secretar\'ia de Ciencia, Humanidades, Tecnolog\'ia e Innovaci\'on,\\ 
Av. Insurgentes Sur 1582. Colonia Cr\'edito Constructor, Del. Benito Ju\'arez, C.P. 03940, Ciudad de M\'exico, M\'exico}
\affiliation[c]{Dual CP Institute of High Energy Physics,\\
C.P. 28045, Colima, M\'exico}

\abstract{
We study the one-loop renormalization of self-interacting massless fields in the $(1,0)\oplus(0,1)$ representation of the Restricted Lorentz Group. We work with a general model that represents the entire class of parity-invariant self-interacting massless theories that can be defined in this representation. It consists of a general free Lagrangian that reproduces the massless limit of three theories previously studied in the literature: the Joos-Weinberg, the Shay-Good/Hammer-McDonald-Pursey, and the Klein-Gordon-like one, as particular cases; along with an interacting Lagrangian containing all the independent dimension-4 parity-invariant self-interactions available in this representation. The model is found to be renormalizable.
}

\makeatletter
\gdef\@fpheader{}
\makeatother

\begin{document}
\maketitle
\flushbottom

\section{Introduction}

According to Wigner's work \cite{WIGNER19899}, one-particle states are defined by their mass $m$ and spin $j$, for $m\neq 0$, or by their helicity $h$, for $m=0$. In quantum field theory (QFT), one-particle states arise from the action of creation Hilbert-space operators on the vacuum state. Likewise, quantum fields are Hilbert-space operators built as linear superpositions of one-particle state creation and annihilation operators with coefficients belonging to a representation of the proper ortochronus Lorentz group $SO^{+}(1,3)$, also known as the Restricted Lorentz Group (RLG) \cite{WeinbergQFT}. The irreducible representations (irreps) of the RLG are labeled by two $SU(2)$ numbers $(a,b)$. 
In general, the $(a,b)$ representation contains multiple spin sectors $j=a+b,\cdots,|a-b|$; among them, one single value has to be picked out, through the addition of restrictions, in the form of equations of motion (EOMs), or as auxiliary equations.
In particular, the irreps $(j,0)$ and $(0,j)$ have one single spin sector, and thus fields
constructed from these irreps have well-defined spin $j$ regardless of the EOM they satisfy. Although there exists an infinite number of RLG irreps and hence an arbitrary number of quantum fields that can be defined out of them, the Standard Model (SM) uses only a few: the scalar $(0,0)$, with $j=0$, for the Higgs; the spinors $(\frac{1}{2},0)$ and $(0,\frac{1}{2})$, with $j=1/2$, for leptons and quarks; and the vector $(\frac{1}{2},\frac{1}{2})$, with $j=1$, for gauge bosons. 
We will refer to them as standard representations, for simplicity.
In view of this, one way to probe physics beyond the SM (BSM) is to survey fields in non-standard representations; in particular, high spin fields ($j>1$) all belong to this class.

The construction of quantum field theories for high spin values or in non-standard RLG representations has been a long-standing problem. Some of the historically most 
popular
approaches addressing it are those due to Dirac-Fierz-Pauli \cite{Dirac:1936tg,Fierz:1939zz,Fierz:1939ix}; Bhabha-Majorana \cite{Bhabha:1945zz,Majorana:1932chs}; and Joos-Weinberg \cite{Joos:1962qq,Weinberg:1964cn,AnySpinII}. Recently, a novel general first-principles formalism for deducing free EOMs for fields in any representation of the RLG was proposed in \cite{Napsuciale:2022usn}. This formalism relies solely on the symmetry group of a free particle: the full Poincaré group (i.e., including parity and time reversal). It asserts that a complete set of commuting observables (CSCO) characterizing its quantum state must include the Hamiltonian, the Casimir operators of the full Poincaré group, the generators of the Cartan subalgebra, and the Hermitian discrete symmetry operators commuting with all the above. Including these last operators is the key feature of this approach; in particular, the parity operator yields a kinematical restriction that turns out to be the EOM.
Thus, it reproduces the conventional EOMs for the standard representations: the Dirac equation for the $(\frac{1}{2},0)\oplus (0,\frac{1}{2})$ and the Proca and Maxwell equations for the $(\frac{1}{2},\frac{1}{2})$ in the massive and massless cases, respectively; in addition, it reveals that, in the latter representation, the gauge symmetry arising in the massless case has its fundamental origin in the parity symmetry. We will refer to this latter approach as the \emph{Poincaré} formalism, for simplicity. Among all these different approaches for constructing fields in non-standard representations, both the Joos-Weinberg and Poincaré formalisms can be applied to a (massive or massless) field belonging to the $(1,0)\oplus(0,1)$ representation of the RLG, and lead to slightly different EOMs that have been studied since decades ago \cite{Joos:1962qq,Weinberg:1964cn,AnySpinII,Shay:1968iq,Hammer,Tucker,Napsuciale:2015kua,Kalb:1974yc,Chizhov:2011,Kaul:1977am,Townsend:1979hd}, with applications ranging from the low energy description of strongly interacting theories like QCD \cite{Ecker:1988te,Chizhov:2011}, to the classical description of the low energy interactions between open strings \cite{Kalb:1974yc} and even dark matter \cite{Hernandez-Arellano:2018sen,Hernandez-Arellano:2019qgd,Hernandez-Arellano:2021bpt}. 
Their corresponding quantum field theories for the massive ($m \neq 0$) cases, however, are not exempt from issues: while the former's propagator has (in addition to the physical pole at $p^{2}=m^{2}$) an unphysical pole at $p^{2}=-m^{2}$ yielding an unconventional QFT, the latter's is ultraviolet(UV)-divergent resulting in a non-renormalizable theory. Interestingly, the propagators for the massless cases do not exhibit either of these issues, and consequently, they could turn out to be conventional and renormalizable QFTs. This observation invites us to study the massless cases, aiming to assess their eventual viability as realistic and fundamental theories for physics BSM. The present article focuses on their renormalization.


All the EOMs for fields in $(1,0)\oplus(0,1)$ are second-order; as a consequence, these fields have mass-dimension one. This fact allows the definition of several dimension-4 interactions with the SM fields as well as dimension-4 self-interactions, opening the door for the exploration of a variety of plausible renormalizable theories. The most straightforward and economical models to start these explorations are precisely those involving only self-interactions, as no additional fields are involved. On the other hand, self-interactions may be required for renormalizing interactions with SM fields. These reasons lead us to study the renormalization of self-interacting massless fields in this representation.

In this work, we study the one-loop renormalization not of a single theory but that of the entire class of parity-invariant self-interacting massless theories for fields in the $(1,0)\oplus(0,1)$ representation. We do so by constructing the most general model compatible with these attributes. This model consists of a free Lagrangian that reproduces the massless limits of three theories previously studied in the literature: the Joos-Weinberg, the Shay-Good/Hammer-McDonald-Pursey, and the Klein-Gordon-like one, as particular cases; along with an interacting Lagrangian containing all the independent dimension-4 parity-invariant self-interactions available in this representation. 

The article is structured as follows: in section \ref{section2}, we review the basics of this representation and deduce the free massive and massless EOMs for fields on it; in section \ref{section3}, we present the model; in section \ref{section4}, the renormalization procedure of the model is performed; and finally, in section \ref{section5}, we end with conclusions.


\section{Fields in \texorpdfstring{$(1,0)\oplus(0,1)$}{(1,0)+(0,1)}}
\label{section2}

A field in the $(1,0)\oplus(0,1)$ representation has six complex components, three coming from each irrep: $(1,0)$ and $(0,1)$. It can be formulated and studied both as a complex six-component column vector $\Psi_{a}$ ($a=1,\cdots,6$) \cite{Joos:1962qq,Weinberg:1964cn,AnySpinII,Shay:1968iq,Hammer,Tucker,Napsuciale:2015kua} or as a complex second-rank antisymmetric Lorentz tensor $B_{\mu\nu}$ $(\mu,\nu=0,\cdots,3)$ \cite{Kalb:1974yc,Chizhov:2011,Kaul:1977am,Townsend:1979hd}, and there exists a map translating between these two languages where each vector index $a$ corresponds to a pair of antisymmetric internal Lorentz indices $\mu\nu$ \cite{Napsuciale:2022usn}. In this work we adopt the former notation.
A covariant basis for operators in this representation was developed in \cite{Selim}. This basis comprises six operators $\{ 1,\chi,S^{\mu\nu},\chi S^{\mu\nu}, M^{\mu\nu}, C^{\mu\nu\rho\sigma}  \}$: the unit operator $1$, the chirality operator $\chi$, two second-rank symmetric tensors $S^{\mu\nu}$ (spin-1 analogous to $\gamma^{\mu}$) and $\chi S^{\mu\nu}$, the Lorentz generators $M^{\mu\nu}$, and a fourth-rank tensor $C^{\mu\nu\rho\sigma}$ with symmetries $        C^{\mu\nu\rho\sigma}=-C^{\nu \mu \rho \sigma}=-C^{\mu\nu\sigma\rho}$ and $C^{\mu\nu\rho\sigma}= C^{\rho\sigma\mu\nu}$, having null traces and satisfying the Bianchi identity. 
The use of this basis makes the mathematical structure of this representation transparent and facilitates the handling of theories constructed from it.

Fields with mass $m\neq 0$ have spin-1, and the  theories studied in the literature can be readily retrieved, along the lines described in \cite{Ecker:1988te,Napsuciale:2013yda}, by writing down the most general parity-invariant free massive Lagrangian in terms of the basis operators and demanding the location of the propagator's pole to be at $p^{2}=m^{2}$.
The most general parity-invariant free massive Lagrangian is
\begin{equation}
    \mathcal{L}
    =
    \partial^{\mu}\bar{\Psi}
    \left(
    \alpha g_{\mu\nu} + \beta S_{\mu\nu} + igM_{\mu\nu}
    \right)
    \partial^{\nu}\Psi
    -
    m^{2}\bar{\Psi}\Psi,
\end{equation}
where $\bar{\Psi}=\Psi^{\dagger}S^{00}$ is the dual field, and $\alpha,\beta$ are arbitrary real-valued coefficients normalized as $|\alpha|+|\beta|=1$ to get proper normalization of the Lagrangian. Its propagator is
\begin{equation}
    i\Gamma_{ab}(p)
    =
    i
    \frac{\left(\alpha p^{2}-m^{2}\right)1_{ab}
    -
    \beta p_{\mu}p_{\nu}(S^{\mu\nu})_{ab}}{\left[(\alpha-\beta)p^{2}-m^{2}\right]
    \left[(\alpha+\beta)p^{2}-m^{2}\right]},
\end{equation}
which has a two-pole structure. To get the physical pole at $p^{2}=m^{2}$ the coefficients must be chosen as $\beta=\pm (\alpha-1)$. The values that have been previously studied in the literature are: i) $\alpha=1,\beta=0$, ii) $\alpha=0,\beta=1$ and iii) $\alpha=\pm \beta=1/2$. For i) we get a Klein-Gordon-like theory, with propagator
\begin{equation}
\label{KGpropagator}
    i\Gamma_{ab}(p)
    =
    i
    \frac{1_{ab}}{p^{2}-m^{2}+i\epsilon};
\end{equation}
this theory and its interactions were studied in \cite{Rangel-Pantoja:2025nkm,Delgado-Acosta:2013nia,Delgado-Acosta:2012dxv,Rivero-Acosta2020}. 
For ii) we get the Joos-Weinberg theory \cite{Joos:1962qq,Weinberg:1964cn}, whose propagator is
\begin{equation}
\label{JWpropagator}
    i\Gamma_{ab}(p)
    =
    i
    \frac{m^{2}1_{ab}
    +
    p_{\mu}p_{\nu}(S^{\mu\nu})_{ab}}{\left[p^{2}+m^{2}\right]
    \left[p^{2}-m^{2}\right]},
\end{equation}
and its interactions were studied in \cite{Eeg:1972us,Eeg:1972uwa,Eeg1973,Chizhov:2011,Avdeev:1993hc}. Finally, for iii) with $\beta=1/2$ we get the Shay-Good/Hammer-McDonald-Pursey\footnote{The EOM was initially proposed, in the column vector language, in \cite{Shay:1968iq,Hammer}. Later obtained, in the antisymmetric Lorentz tensor language, in \cite{Kalb:1974yc,Ecker:1988te}. Its canonical quantization was performed in \cite{Napsuciale:2015kua}. Also, it was deduced from the Poincaré formalism and shown the equivalence of both formulations in \cite{Napsuciale:2022usn}.} 
theory \cite{Shay:1968iq,Hammer},
whose propagator is
\begin{equation}
\label{SGpropagator}
    i\Gamma_{ab}(p)
    =
    i
    \frac{\left(-p^{2}+2m^{2}\right)1_{ab}
    +
    p_{\mu}p_{\nu}(S^{\mu\nu})_{ab}}{2m^{2}
    \left[p^{2}-m^{2}+i\epsilon\right]},
\end{equation}
and its interactions were studied in \cite{Prabhakaran:1973pr,Napsuciale:2015kua,Chizhov:2011}. This latter theory is the one arising from the Poincaré formalism in \cite{Napsuciale:2022usn}, where it was found that the case $\beta=1/2$ corresponds to the positive-parity field construction, while the case $\beta=-1/2$ corresponds to the negative-parity one; it is also shown there that, when written in the tensor language, it becomes the massive ($m\neq 0$) Kalb-Ramond theory \cite{Kalb:1974yc}.
Inspection of propagators \eqref{KGpropagator}--\eqref{SGpropagator} allows us to gain insight into the features of the theories we are dealing with.
For theory i) we have a conventional Klein-Gordon-like propagator \eqref{KGpropagator}, that is $1/p^{2}$-order and a single pole at $p^{2}=m^{2}$; in fact, the one-loop renormalization of its dimension-4 self-interactions and electrodynamics was demonstrated in \cite{Rivero-Acosta2020}.
For theory ii) we have propagator \eqref{JWpropagator}, which is $1/p^{2}$-order and less, but has an additional and awkward unphysical pole located at $p^{2}=-m^{2}$.
For theory iii) we have propagator \eqref{SGpropagator}, which has a single pole at $p^{2}=m^{2}$ but is zero-order and less in $p^{2}$, implying that it renders amplitude integrals divergent in the UV region ($p^{2}\gg m^{2}$) and thus leads to a non-renormalizable theory (it suffers from the same problem as the Proca massive vector field).

In contrast, the massless theories do not have any of those problems. The most general parity-invariant free massless Lagrangian is
\begin{equation}
\label{massless Lagrangian}
    \mathcal{L}
    =
    \partial^{\mu}\bar{\Psi}
    \left(
    \alpha g_{\mu\nu} + \beta S_{\mu\nu} + igM_{\mu\nu}
    \right)
    \partial^{\nu}\Psi,
\end{equation}
and its propagator becomes
\begin{equation}
\label{massless propagator}
i\Gamma_{ab}(p)
=
\frac{i}{\alpha^{2}-\beta^{2}}
\frac{ \alpha p^{2} 1_{ab} - \beta  p_{\mu}p_{\nu} (S^{\mu\nu})_{ab} }{p^{4}+i\epsilon},
\end{equation}
which is $1/p^{2}$-order, making it well-behaved in the UV region, and no longer has an unphysical pole. Notice, however, that it's undefined for $|\alpha|=|\beta|$, which is precisely the massless case for theory iii). As explained in \cite{Napsuciale:2022usn}, this is due to the emergence of a gauge symmetry, implemented as $\Psi \rightarrow \Psi'=\Psi + (\alpha g_{\mu\nu}-\beta S_{\mu\nu}) \partial^{\mu} \partial^{\nu} \Phi$ (which in the tensor language becomes the gauge symmetry of the massless Kalb-Ramond field \cite{Kalb:1974yc}), and, to get a well-defined propagator, it's necessary to add the gauge-fixing term 
\begin{equation}
    \mathcal{L}_{\text{gf}}
    =
    \frac{1}{\xi}
    \partial^{\mu}\bar{\Psi}
    \left(
    \alpha g_{\mu\nu} - \beta S_{\mu\nu}
    \right)
    \partial^{\nu}\Psi,
    \qquad
    \qquad
    \text{for }
    \alpha=\pm \beta = \frac{1}{2},
\end{equation}
where $\xi$ is the gauge parameter. With this additional term, the Lagrangian for the massless theory iii) becomes
\begin{equation}
\label{Lagrangianiii}
    \mathcal{L}
    =
    \partial^{\mu}\bar{\Psi}
    \left[
    \frac{1}{2}
    \left(1+\frac{1}{\xi}\right)g_{\mu\nu} 
    \pm 
    \frac{1}{2}\left(1-\frac{1}{\xi}\right) S_{\mu\nu} 
    +
    igM_{\mu\nu}
    \right]
    \partial^{\nu}\Psi,
\end{equation}
which has the form of \eqref{massless Lagrangian} with $\alpha=(1/2)(1+1/\xi)$ and $\beta=\pm (1/2)(1-1/\xi)$ satisfying now $|\alpha|\neq |\beta|$ for finite $\xi$, and hence its propagator is still a particular case of \eqref{massless propagator}.
From these observations, we can conclude that the propagator \eqref{massless propagator} is the most general one for massless fields in this representation and doesn't suffer from the problems that the propagators of theories with mass have.

At this point, it's important to note that the gauge symmetry arising in the massless theory iii) has led to interpreting this field not as a matter field but as a gauge one. This distinction is further explained in \cite{Chizhov:2011}, using tensor language, where the matter and gauge theories presented there correspond to ii) and iii), respectively. Nevertheless, the study of theory iii), with $m\neq 0$, as a matter field has yielded interesting results when proposed as dark matter and gave rise to a new candidate called \textit{tensor dark matter} (TDM) \cite{Hernandez-Arellano:2018sen,Hernandez-Arellano:2019qgd,Hernandez-Arellano:2021bpt}.

From the Lagrangian \eqref{massless Lagrangian}, we see that the field's mass dimension is one. This fact enables for the formulation of dimension-4 self-interaction operators and makes possible the construction of power-counting renormalizable self-interacting theories. In the next section, we construct the general model we work with.





\section{The model}
\label{section3}

In order to address the renormalization of the entire class of parity-invariant self-interacting massless theories, we have to construct the most general model compatible with these attributes. As we saw above, the most general free Lagrangian is given by \eqref{massless Lagrangian}, which has the massless limits of theories i), ii), and iii) as particular cases. The term involving $ ig M_{\mu\nu}$ becomes relevant for gauge interactions, see for example \cite{Rivero-Acosta2020}; however, for self-interactions it's just an insignificant boundary term, and so we will skip it in the following. The most general interaction Lagrangian comprises all the independent dimension-4 parity-invariant self-interactions; these are quartic interactions of the form $(\bar{\Psi}A\Psi) (\bar{\Psi}B\Psi)$, where $A$ and $B$ are  operators of the covariant basis $\{ 1,\chi,S^{\mu\nu},\chi S^{\mu\nu}, M^{\mu\nu}, C^{\mu\nu\rho\sigma}  \}$. 
By deducing the corresponding Fierz identities it is shown that, among the parity-invariant, the independent ones are just four: $(\bar{\Psi}\Psi)^{2},(\bar{\Psi}\chi\Psi)^{2},(\bar{\Psi}S^{\mu\nu}\Psi)^{2},(\bar{\Psi}M^{\mu\nu}\Psi)^{2}$. 
Thus, the model is described by the Lagrangian
\begin{equation}
\label{LagrangianB}
    \mathcal{L}
    =
    \partial^{\mu} \bar{\Psi}
    \left(
    \alpha
    g_{\mu\nu}
    +
    \beta
    S_{\mu\nu}
    \right)
    \partial^{\nu}
    \Psi
    +
    \frac{\lambda_{1}}{2}
    (\bar{\Psi}\Psi)^{2}    
    +
    \frac{\lambda_{2}}{2}
    (\bar{\Psi} \chi \Psi)^{2}   
    +
    \frac{\lambda_{3}}{2}
    (\bar{\Psi} S^{\mu\nu} \Psi)^{2}  
    +
    \frac{\lambda_{4}}{2}
    (\bar{\Psi} M^{\mu\nu} \Psi)^{2},
\end{equation}
where $\lambda_{1},\lambda_{2},\lambda_{3},\lambda_{4}$ are coupling parameters.
Here it's important to notice that, even though the kinetic term in \eqref{LagrangianB} has massless theory iii), in \eqref{Lagrangianiii}, as a particular case and so its propagator, the introduced self-interactions explicitly break its gauge symmetry. This may bring inconsistencies in the propagating degrees of freedom of its interacting theory, and to avoid this possible issue, we restrict to the study of theories i) and ii) from now on. The Feynman rules for \eqref{Lagrangianiii} are
\begin{align}
\label{FR1}
\begin{tikzpicture}[baseline=-\the\dimexpr\fontdimen22\textfont2\relax]
\begin{feynman}[inline=(o)]
\vertex (o) ;
\vertex at ($(o) + (-1cm, 0cm)$) (i1);
\vertex at ($(o) + (1cm, 0cm)$) (i2);
\vertex at ($(o) + (0cm, 1cm)$) (i3);
\vertex at ($(o) + (0cm, -1cm)$) (i4);
\vertex at ($(o) + (-1cm, 1cm)$) (v1);
\vertex at ($(o) + (1cm, 1cm)$) (v2);
\vertex at ($(o) + (-1cm, -1cm)$) (v3);
\vertex at ($(o) + (1cm, -1cm)$) (v4);
    \diagram* {
    (i1) -- [double,with arrow=0.5, edge label=$p$] (i2),
    };
    \filldraw (o) node {$a$\hspace{2.3cm}$b$};
\end{feynman}
\end{tikzpicture}
\equiv
&
i\Gamma_{ab}(p)
=
\frac{i}{\alpha^{2}-\beta^{2}}
\frac{ \alpha p^{2} 1_{ab} - \beta  p_{\mu}p_{\nu} (S^{\mu\nu})_{ab} }{p^{4}+i\epsilon},
\\
\label{FR2}
\begin{tikzpicture}[baseline=-\the\dimexpr\fontdimen22\textfont2\relax]
\begin{feynman}[inline=(o)]
    \diagram* {
    (v3) -- [double,with arrow=0.5, edge label=${b,p_2}$] (o)
    --[double,with arrow=0.5, edge label=${d,p_4}$] (v4),
    (v1) -- [double,with arrow=0.5, edge label'=${a,p_1}$] (o)
    --[double,with arrow=0.5, edge label'=${c,p_3}$] (v2)
    };
    \filldraw[black] (o) circle (1.5pt);
\end{feynman}
\end{tikzpicture}
\equiv
&
iV_{abcd}
=
\begin{array}{l}
+i\lambda_{1}
\left[
1_{ab}1_{cd}
+
1_{ad}1_{cb}
\right]
+i\lambda_{2}
\left[
\chi_{ab}\chi_{cd}
+
\chi_{ad}\chi_{cb}
\right]
\\
+i\lambda_{3}
\left[
(S^{\mu\nu})_{ab}(S_{\mu\nu})_{cd}
+
(S^{\mu\nu})_{ad}(S_{\mu\nu})_{cb}
\right]
\\
+i\lambda_{4}
\left[
(M^{\mu\nu})_{ab}(M_{\mu\nu})_{cd}
+
(M^{\mu\nu})_{ad}(M_{\mu\nu})_{cb}
\right],
\end{array}
\end{align}
where latin subindices $a,b,c,d$ are column-vector ones $a,b,c,d=1,\cdots,6$. We address the renormalization procedure in the next section.

\section{Renormalization}
\label{section4}

In this section, we conduct the one-loop renormalization of \eqref{LagrangianB} in the minimal subtraction scheme (MS) using dimensional regularization (DR) as the UV regulator, within the context of renormalized perturbation theory. 
From the Feynman rules \eqref{FR1} and \eqref{FR2}, the superficial degree of divergence $\mathcal{D}$ of a diagram with $N$ external legs is given by $\mathcal{D}=4-N$ and so there are only two non-trivial potentially divergent diagrams: those with $N=2$ and $N=4$, which are the self-energy and the vertex, respectively
\begin{align}
\label{DivergentDiagrams}
1)
\hspace{0.1cm}
\begin{tikzpicture}[baseline=-\the\dimexpr\fontdimen22\textfont2\relax]
\begin{feynman}[inline=(o)]
\vertex at ($(o) + (-0.5cm, 0cm)$) (i1/2);
\vertex at ($(o) + (0.5cm, 0cm)$) (i2/2);
    \diagram* {
    (i1) -- [double] 
    (o) -- [double] (i2)
    };
    \draw[black, fill=gray] (i2/2) node[circle, inner sep=1pt] node {} arc(0:360:0.5);
\end{feynman}
\end{tikzpicture},
&
&
&
2)
\hspace{0.1cm}
\begin{tikzpicture}[baseline=-\the\dimexpr\fontdimen22\textfont2\relax]
\begin{feynman}[inline=(o)]
    \diagram* {
    (v1) -- [double]  (o)
    -- [double] (v2),
    (v3) -- [double]  (o)
    -- [double] (v4)
    };
    \draw[black, fill=gray] (i2/2) node[circle, inner sep=1pt] node {} arc(0:360:0.5);
\end{feynman}
\end{tikzpicture}.
\end{align}

To implement DR, we first have to extend the basis $\{ 1,\chi,S^{\mu\nu},\chi S^{\mu\nu}, M^{\mu\nu}, C^{\mu\nu\rho\sigma}  \}$ to dimension $D$. Initially, this task seems challenging because its Lie and Jordan algebras are very complicated (much more than those of the Dirac basis), as can be explicitly seen in \cite{Selim}; for example, the anticommutator $\{ S^{\mu\nu}, S^{\rho \sigma} \}$ is given as
\begin{equation}
\label{clifford}
\{ S_{\mu\nu}, S_{\rho\sigma} \} =
\frac{4}{3} \left( 
g_{\mu\rho} g_{\nu\sigma} + g_{\mu\sigma} g_{\nu\rho} 
- \frac{1}{2} g_{\mu\nu} g_{\rho\sigma} 
\right)
- \frac{1}{6} \left( 
C_{\mu\rho\nu\sigma} + C_{\mu\sigma\nu\rho} 
\right),
\end{equation}
depending not only on $g_{\mu\nu}$, but also on $C_{\mu\nu\rho\sigma}$. Thus, if we define the $D$-dimensional $S^{\mu\nu}$ by demanding fulfillment of the algebra \eqref{clifford} (as is done for the Dirac gamma matrices in the context of DR, see for example \cite{Peskin:1995ev}), all the results coming from it will involve the $D$-dimensional $C_{\mu\nu\rho\sigma}$ as well, and so this approach gets hard to follow. Fortunately, there exists an alternative. In the tensor language all of the basis operators are expressed in terms of the metric tensor $g_{\mu\nu}$ and the Levi-Civita symbol $\varepsilon_{\mu\nu\rho\sigma}$, as \cite{Napsuciale:2013yda}
\begin{align}
\label{1Ap}
    1_{\alpha \beta \gamma \delta}
    =&
    \frac{1}{2}
    \left(
    g_{\alpha \gamma} g_{\beta \delta}
    -
    g_{\alpha \delta} g_{\beta \gamma}
    \right),
    \\
\label{ChiAp}
    \chi_{\alpha \beta \gamma \delta}
    =&
    \frac{i}{2} \varepsilon_{\alpha \beta \gamma \delta},
\\    
\label{SAp}
(S_{\mu\nu})_{\alpha \beta \gamma \delta}
=&
g_{\mu\nu}
1_{\alpha \beta \gamma \delta}
-
g_{\mu\gamma}
1_{\alpha \beta \nu \delta}
-
g_{\mu\delta}
1_{\alpha \beta \gamma \nu}
-
g_{\nu\gamma}
1_{\alpha \beta \mu \delta}
-
g_{\nu\delta}
1_{\alpha \beta \gamma \mu},
\\
\label{MAp}
    (M_{\mu \nu})_{\alpha \beta \gamma \delta}
    =&
    -i(g_{\mu \gamma} 1_{\alpha \beta \nu \delta} 
    +
    g_{\mu \delta} 1_{\alpha \beta \gamma \nu}
    -
    g_{\gamma \nu} 1_{\alpha \beta \mu \delta}
    -
    g_{\delta \nu} 1_{\alpha \beta \gamma \mu}
    ),
    \\
\label{CAp}
    (C_{\mu\nu \rho \sigma})_{\alpha \beta \gamma \delta}
    =&
    32 1_{\alpha \gamma \mu \nu}
    1_{\delta \beta \rho \sigma}
    -
    32 1_{\alpha \gamma \rho \sigma}
    1_{\beta \delta \mu \nu}
    +
    6 g_{\alpha \gamma} X_{\beta \delta \rho \sigma \mu \nu}
    +
    6 g_{\beta \delta}
    X_{\alpha \gamma \rho \sigma \mu \nu}
    \\
    \nonumber
    &+8 1_{\alpha\beta\delta \gamma}
    1_{\rho \sigma \mu \nu}
    +16 1_{\alpha \beta \rho \sigma} 1_{\delta \gamma \mu \nu}
    +16 1_{\alpha \beta \mu \nu}
    1_{\delta \gamma \rho \sigma}
    -16 1_{\alpha \delta \rho \sigma} 1_{\beta \gamma \mu \nu}
    \\
    \nonumber
    &
    -16 1_{\alpha \delta \mu \nu} 1_{\beta \gamma \rho \sigma},
\end{align}
where
\begin{equation}
    X_{\alpha \gamma \rho \sigma \mu \nu}
    =
    2 g_{\alpha \rho } 1_{\mu \nu \gamma \sigma}
    -2 g_{\alpha \sigma} 1_{\mu \nu \gamma \rho}
    -2 g_{\alpha \mu} 1_{\nu \gamma \rho \sigma}
    +2 g_{\alpha \nu} 1_{\mu \gamma \rho \sigma}.
\end{equation}
Thus, this language provides a straightforward way to construct the $D$-dimensional extension of the basis: simply replace the $4$-dimensional metric tensor $g_{\mu\nu}$ with the $D$-dimensional one and follow a prescription for the Levi-Civita symbol $\varepsilon_{\mu\nu\rho\sigma}$. In fact, for the self-interacting theory \eqref{LagrangianB}, it's not even necessary to worry about $\varepsilon_{\mu\nu\rho\sigma}$ because the term involving it, in the Feynman rule \eqref{FR2}, can be written in terms of $g_{\mu\nu}$ as well by using the identity
\begin{equation}
\chi_{ab}\chi_{cd}
\overset{\cdot}{=}
\chi_{a_{1} a_{2} b_{1} b_{2}}
\chi_{c_{1} c_{2} d_{1} d_{2}}
=
\frac{i^2}{4}
    \varepsilon_{a_{1} a_{2} b_{1} b_{2}}
    \varepsilon_{c_{1} c_{2} d_{1} d_{2}}
    =
\frac{1}{4}
    \text{Det}
\begin{pmatrix}
 g_{a_{1} c_{1}} & g_{a_{1} c_{2}} & g_{a_{1} d_{1}} & g_{a_{1} d_{2}} \\
 g_{a_{2} c_{1}} & g_{a_{2} c_{2}} & g_{a_{2} d_{1}} & g_{a_{2} d_{2}} \\
 g_{b_{1} c_{1}} & g_{b_{1} c_{2}} & g_{b_{1} d_{1}} & g_{b_{1} d_{2}} \\
 g_{b_{2} c_{1}} & g_{b_{2} c_{2}} & g_{b_{2} d_{1}} & g_{b_{2} d_{2}} \\
\end{pmatrix},
\end{equation}
where $\{(a_{1},a_{2}),(b_{1},b_{2}),(c_{1},c_{2}),(d_{1},d_{2})\}$ are the internal Lorentz indices corresponding to the column-vector ones $\{a,b,c,d\}$, respectively; and hence, in this case, it all comes down to the metric tensor. 

That being explained, we proceed to show the explicit calculations. In the following subsections, we derive the counterterms, compute the amplitudes for the divergent diagrams \eqref{DivergentDiagrams}, determine the values of the counterterms, evaluate the beta functions, and find their fixed points.

\subsection{Counterterms}

The Lagrangian in \eqref{LagrangianB} is explicitly written in terms of the bare field and the bare coupling parameters, which we denote with a subscript $0$ from now on: $\Psi_{0}$ and $ \lambda_{X,0}$. The replacement of the bare field $\Psi_{0}$ in favor of the physical one $\Psi$ through the relation $\Psi=Z^{-\frac{1}{2}} \Psi_{0}$ gives rise to the split of the Lagrangian into physical terms $\mathcal{L}_{\text{phys}}$ and counterterms $\mathcal{L}_{\text{ct}}$ as $\mathcal{L} = \mathcal{L}_{\text{phys}} + \mathcal{L}_{\text{ct}}$, reading
\begin{equation}
\begin{aligned}
    \mathcal{L}
=&
    \partial^{\mu} \bar{\Psi}
    \left(
    \alpha
    g_{\mu\nu}
    +
    \beta
    S_{\mu\nu}
    \right)
    \partial^{\nu}
    \Psi
    +
    \frac{\lambda_{1}}{2}
    (\bar{\Psi}\Psi)^{2}    
    +
    \frac{\lambda_{2} }{2}
    (\bar{\Psi} \chi \Psi)^{2} 
    +
    \frac{\lambda_{3}}{2}
    (\bar{\Psi} S^{\mu\nu} \Psi)^{2}  
    +
    \frac{\lambda_{4} }{2}
    (\bar{\Psi} M^{\mu\nu} \Psi)^{2}
\\
&
+
    \textcolor{purple}{
    \delta_{Z}}
    \partial^{\mu} \bar{\Psi}
    \left(
    \alpha
    g_{\mu\nu}
    +
    \beta
    S_{\mu\nu}
    \right)
    \partial^{\nu}
    \Psi
    +
\textcolor{purple}{
\delta_{Z_{1}}}
    \frac{\lambda_{1}}{2}
    (\bar{\Psi}\Psi)^{2}    
    +
\textcolor{purple}{
\delta_{Z_{2}}}
    \frac{\lambda_{2} }{2}
    (\bar{\Psi} \chi \Psi)^{2}  
    +
\textcolor{purple}{
\delta_{Z_{3}}}
    \frac{\lambda_{3}}{2}
    (\bar{\Psi} S^{\mu\nu} \Psi)^{2}  
\\
&
    +
\textcolor{purple}{
\delta_{Z_{4}}}
    \frac{\lambda_{4} }{2}
    (\bar{\Psi} M^{\mu\nu} \Psi)^{2},
\end{aligned}
\end{equation}
being the terms in the first line the physical ones, and those in the second and third lines the counterterms;
the $\textcolor{purple}{\delta_{X}}$ are defined as
\begin{align}
\label{deltas}
    \textcolor{purple}{\delta_{Z}=}&\textcolor{purple}{Z-1},
    &
    \textcolor{purple}{\delta_{Z_{1}}=}
    &
    \textcolor{purple}{
    Z_{1}
    -1},
    &
    \textcolor{purple}{\delta_{Z_{2}}=}
    &
    \textcolor{purple}{
    Z_{2}
    -1},
    &
    \textcolor{purple}{\delta_{Z_{3}}=}
    &
    \textcolor{purple}{
    Z_{3}
    -1},
    &
    \textcolor{purple}{\delta_{Z_{4}}=}
    &
    \textcolor{purple}{
    Z_{4}
    -1},
\end{align}
with the renormalization constants $Z_{i}$ given by
\begin{align}
\label{lambdas}
Z_{1}
=&
\frac{\lambda_{1,0}}{\lambda_{1}}Z^{2},
&
Z_{2}
=&
\frac{\lambda_{2,0}}{\lambda_{2}}Z^{2},
&
Z_{3}
=&
\frac{\lambda_{3,0}}{\lambda_{3}}Z^{2},
&
Z_{4}
=&
\frac{\lambda_{4,0}}{\lambda_{4}}Z^{2}.
\end{align}
Notice that, to keep the physical coupling parameters $\lambda_{X}$ dimensionless in DR with $D=4-2\epsilon$, we have to scale them with an arbitrary energy scale $\mu$ according to
\begin{align}
\label{scale}
    \lambda_{1}\rightarrow & \mu^{2\epsilon}\lambda_{1},
    &
    \lambda_{2}\rightarrow & \mu^{2\epsilon}\lambda_{2},
    &
    \lambda_{3}\rightarrow & \mu^{2\epsilon}\lambda_{3},
    &
    \lambda_{4}\rightarrow & \mu^{2\epsilon}\lambda_{4}.
\end{align}
The Feynman rules for the physical terms are
exactly the same as those in \eqref{FR1}, being now $\lambda_{X}$ the actual physical coupling parameters instead of the bare ones $\lambda_{X,0}$;
in addition, the Feynman rules for the counterterms are
\begin{align}
\label{FR5}
\begin{tikzpicture}[baseline=-\the\dimexpr\fontdimen22\textfont2\relax]
\begin{feynman}[inline=(o)]
    \diagram* {
    (i1) -- [double,with arrow=0.5, edge label=$p$] (o)-- [double,with arrow=0.5, edge label=$p$] (i2),
    };
    \filldraw (o) node {$a$\hspace{2.3cm}$b$};
    \draw[fill=white,cross] 
    (o) circle (0.15);
\end{feynman}
\end{tikzpicture}
\equiv
&
i\delta\Gamma_{ab}(p)
=
i
\textcolor{purple}{\delta_{Z}}
\left[
\alpha p^{2} 1_{ab}
+ 
\beta p_{\mu}p_{\nu} (S^{\mu\nu})_{ab}
\right],
\\
\label{FR6}
\begin{tikzpicture}[baseline=-\the\dimexpr\fontdimen22\textfont2\relax]
\begin{feynman}[inline=(o)]
    \diagram* {
    (v3) -- [double,with arrow=0.5, edge label=${b,p_2}$] (o)
    --[double,with arrow=0.5, edge label=${d,p_4}$] (v4),
    (v1) -- [double,with arrow=0.5, edge label'=${a,p_1}$] (o)
    --[double,with arrow=0.5, edge label'=${c,p_3}$] (v2)
    };
    \draw[fill=white,cross] 
    (o) circle (0.15);
\end{feynman}
\end{tikzpicture}
\equiv
&
i\delta V_{abcd}
=
\begin{array}{l}
+i\lambda_{1}
\textcolor{purple}{\delta_{Z_{1}}}
\left[
1_{ab}1_{cd}
+
1_{ad}1_{cb}
\right]
+i\lambda_{2}
\textcolor{purple}{\delta_{Z_{2}}}
\left[
\chi_{ab}\chi_{cd}
+
\chi_{ad}\chi_{cb}
\right]
\\
+i\lambda_{3}
\textcolor{purple}{\delta_{Z_{3}}}
\left[
(S^{\mu\nu})_{ab}(S_{\mu\nu})_{cd}
+
(S^{\mu\nu})_{ad}(S_{\mu\nu})_{cb}
\right]
\\
+i\lambda_{4}
\textcolor{purple}{\delta_{Z_{4}}}
\left[
(M^{\mu\nu})_{ab}(M_{\mu\nu})_{cd}
+
(M^{\mu\nu})_{ad}(M_{\mu\nu})_{cb}
\right].
\end{array}
\end{align}
With the Feynman rules for physical terms \eqref{FR1}, \eqref{FR2} and counterterms \eqref{FR5},\eqref{FR6} in hand, we now move on to the regularization of the divergent diagrams' amplitudes, i.e. the self-energy and the vertex,  and the determination of the values for the counterterms.

\subsection{Self-energy}
The diagrams contributing to the self-energy at one-loop $i\Sigma(p)$ are
\begin{align}
i\Sigma_{ad}(p)
=&
\begin{tikzpicture}[baseline=-\the\dimexpr\fontdimen22\textfont2\relax]
\begin{feynman}[inline=(o)]
\vertex (o) ;
\vertex at ($(o) + (-1.75cm, 0cm)$) (i1);
\vertex at ($(o) + (-0.75cm, 0cm)$) (i2);
\vertex at ($(o) + (0.75cm, 0cm)$) (i3);
\vertex at ($(o) + (1.75cm, 0cm)$) (i4);
\vertex at ($(o) + (0cm, -0.75cm)$) (v1);
\vertex at ($(o) + (-0.5cm, 0cm)$) (x);
\vertex at ($(o) + (0cm,0.75cm) + (135:0.75)$) (x2);
    \diagram* {
    (i1) -- [double,with arrow=0.5, edge label={$p$}] (i2),
    (i2) -- [double] (i3),
    (i3) -- [double,with arrow=0.5, edge label={$p$}] (i4),
    };
    \node[below] (3) at (0,1.5) ;
    \draw[double, with reversed arrow=0.5] (o)  arc(-90:270:0.75) node[pos=0.5, below] {$l$};
    \filldraw[black] (o) circle (1.5pt);
    \filldraw (o) node {$a$\hspace{3.75cm}$d$};
\end{feynman}
\end{tikzpicture}
+
\begin{tikzpicture}[baseline=-\the\dimexpr\fontdimen22\textfont2\relax]
\begin{feynman}[inline=(o)]
    \diagram* {
    (i1) -- [double,with arrow=0.5, edge label={$p$}] (x),
    (x) -- [double,,with arrow=0.5, edge label={$p$}] (i3),
    };
    \draw[fill=white,cross] 
    (x) circle (0.15);
    \filldraw (x) node {$a$\hspace{2.75cm}$d$};
\end{feynman}
\end{tikzpicture},
\end{align}
and its amplitude in DR is
\begin{equation}
\begin{split}
i\Sigma_{ad}(p)
=&
\mu^{2\epsilon}
\int \frac{d^{D}l}{(2\pi)^{D}}
iV_{abcd}
i\Gamma_{bc}(l)
+
i\delta \Gamma_{ad}(p).
\end{split}
\end{equation}
By adding a fictitious mass $m$ to the propagator, setting $D=4-2\epsilon$, and taking the limit $\epsilon \rightarrow 0$, we find that the UV-divergent part is
\begin{equation}
\begin{split}
\text{Div}
\left[
i\Sigma_{ad}(p)
\right]
=&
\frac{-i\alpha m^{2} (7\lambda_{1} +\lambda_{2}+12 \lambda_{3}+8 \lambda_{4}) }{8\pi^{2}(\alpha^2-\beta^2) \epsilon}
1_{ad}
+
i 
\textcolor{purple}{\delta_{Z}}
\left[
\alpha 
p^2
1_{ad}
+
\beta 
p_{\mu}p_{\nu}(S^{\mu\nu})_{ad}
\right],
\end{split}
\end{equation}
and thus, after taking $m \rightarrow 0$, the divergence disappears and so the counterterm $\textcolor{purple}{\delta_{Z}}$, in the MS scheme, must be chosen as
\begin{equation}
\label{deltaZ=0}
    \textcolor{purple}{\delta_{Z}}=0.
\end{equation}

\subsection{Vertex}

The diagrams contributing to the vertex at one-loop $i\Lambda(p_{1},p_{2},p_{3})$ are
\begin{align}
&i\Lambda_{abcd}(p_{1},p_{2},p_{3})=
\\
&
\begin{tikzpicture}[baseline=-\the\dimexpr\fontdimen22\textfont2\relax]
\begin{feynman}[inline=(o)]
\vertex (o) ;
\vertex at ($(o) + (-1cm, 0cm)$) (i1);
\vertex at ($(o) + (1cm, 0cm)$) (i2);
\vertex at ($(o) + (0cm, 1cm)$) (i3);
\vertex at ($(o) + (0cm, -1cm)$) (i4);
\vertex at ($(o) + (-1cm, 1cm)$) (v1);
\vertex at ($(o) + (1cm, 1cm)$) (v2);
\vertex at ($(o) + (-1cm, -1cm)$) (v3);
\vertex at ($(o) + (1cm, -1cm)$) (v4);
\vertex at ($(o) + (-0.5cm, 0cm)$) (o-1);
\vertex at ($(o) + (0.5cm, 0cm)$) (o+1);
    \diagram* {
    (v1) -- [double,with arrow=0.5, edge label'=${a,p_1}$] (o-1)
    -- [double,with reversed arrow=0.5, edge label'=${b,p_2}$] (v3),
    (v2) -- [double,with reversed arrow=0.5, edge label=${c,p_3}$]  (o+1)
    -- [double,with arrow=0.5, edge label=${d,p_4}$] (v4)
    };
    \draw[double] (i2/2) node[circle, inner sep=1pt] node {} arc(0:360:0.5);
    \filldraw[black] (o-1) circle (1.5pt);
    \filldraw[black] (o+1) circle (1.5pt);
\end{feynman}
\end{tikzpicture}
+
\begin{tikzpicture}[baseline=-\the\dimexpr\fontdimen22\textfont2\relax]
\begin{feynman}[inline=(o)]
\vertex at ($(o) + (0cm, -0.5cm)$) (o-1y);
\vertex at ($(o) + (0cm, 0.5cm)$) (o+1y);
    \diagram* {
    (v1) -- [double,with arrow=0.5, edge label'=${a,p_1}$] (o+1y)
    -- [double,with arrow=0.5, edge label'=${c,p_3}$] (v2),
    (v3) -- [double,with arrow=0.5, edge label=${b,p_2}$] (o-1y)
    -- [double,with arrow=0.5, edge label=${d,p_4}$] (v4)
    };
    \draw[double] (i2/2) node[circle, inner sep=1pt] node {} arc(0:360:0.5);
    \filldraw[black] (o-1y) circle (1.5pt);
    \filldraw[black] (o+1y) circle (1.5pt);
\end{feynman}
\end{tikzpicture}
+
\begin{tikzpicture}[baseline=-\the\dimexpr\fontdimen22\textfont2\relax]
\begin{feynman}[inline=(o)]
    \diagram* {
    (v1) -- [double,with arrow=0.5, edge label'=${a,p_1}$]  (o+1y)
    -- [double, edge label=${c,p_3}$, quarter left] (v4),
    (v3) -- [double,with arrow=0.5, edge label=${b,p_2}$] (o-1y)
    -- [double, edge label'=${d,p_4}$
    , quarter right
    ] (v2) 
    };
    \draw[double] (i2/2) node[circle, inner sep=1pt] node {} arc(0:360:0.5);
    \filldraw[black] (o-1y) circle (1.5pt);
    \filldraw[black] (o+1y) circle (1.5pt);
\end{feynman}
\end{tikzpicture}
+
\begin{tikzpicture}[baseline=-\the\dimexpr\fontdimen22\textfont2\relax]
\begin{feynman}[inline=(o)]
    \diagram* {
    (v1) -- [double,with arrow=0.5, edge label'=${a,p_1}$]  (o)
    -- [double, with arrow=0.5, edge label=${d,p_4}$] (v4),
    (v3) -- [double, with arrow=0.5, edge label=${b,p_2}$] (o)
    -- [double, with arrow=0.5, edge label'=${c,p_3}$] (v2),
    };
    \draw[fill=white,cross] 
    (o) circle (0.15);
\end{feynman}
\end{tikzpicture}
\nonumber
\end{align}
and its amplitude in DR is
\begin{equation}
\begin{aligned}
    i\Lambda_{abcd}(p_{1},p_{2},p_{3})
    =&
    \mu^{2\epsilon}
    \int
    \frac{d^{D}l}{(2\pi)^{D}}
    \left[
    iV_{abrn}
    i\Gamma_{nm}(l+p_{1}+p_{2})
    iV_{mdcs}
    i\Gamma_{sr}(l)
    \right.
    \\
    &
    \hspace{1.25cm}
    +
    iV_{arcn}
    i\Gamma_{nm}(l+p_{1}-p_{3})
    iV_{mdsb}
    i\Gamma_{sr}(l)
    \\
    &
    \hspace{1.25cm}
    \left.
    +
    iV_{adnr}
    i\Gamma_{nm}(l+p_{1}-p_{4})
    iV_{mcbs}
    i\Gamma_{sr}(l)
    \right]
+i\delta V_{abcd}
\end{aligned}
\end{equation}
Setting $D=4-2\epsilon$ and taking the limit $\epsilon \rightarrow 0$, we find that the UV-divergent part is
\begin{equation}
\label{VertexDivergence}
\begin{split}
&
\text{Div}
\left[
i\Lambda_{abcd}(p_{1},p_{2},p_{3})
\right]
=
\frac{-i}{(4\pi)^{2}(\alpha^{2}-\beta^{2})^{2}\epsilon}
\\
&
\times
\left\{
\left(
\alpha^{2}
c_{1}^{1}
+
\frac{\beta^{2}}{12}
c_{1}^{S}
\right)
\left[
1_{ab}
1_{cd}
+
1_{ad}
1_{cb}
\right]
+
\left(
\alpha^{2}
c_{\chi}^{1}
+
\frac{\beta^{2}}{12}
c_{\chi}^{S}
\right)
\left[
\chi_{ab}
\chi_{cd}
+
\chi_{ad}
\chi_{cb}
\right]
\right.
\\
&
+
\left(
\alpha^{2}
c_{S}^{1}
+
\frac{\beta^{2}}{12}
c_{S}^{S}
\right)
\left[
    (S^{\mu\nu})_{ab}
    (S_{\mu\nu})_{cd}
    +
    (S^{\mu\nu})_{ad}
    (S_{\mu\nu})_{cb}
\right]
\\
&
\left.
+
\left(
\alpha^{2}
c_{M}^{1}
+
\frac{\beta^{2}}{12}
c_{M}^{S}
\right)
\left[
    (M^{\mu\nu})_{ab}
    (M_{\mu\nu})_{cd}
    +
    (M^{\mu\nu})_{ad}
    (M_{\mu\nu})_{cb}
\right]
\right\}
\\
&
+i\lambda_{1}
\textcolor{purple}{\delta_{Z_{1}}}
\left[
1_{ab}
1_{cd}
+
1_{ad}
1_{cb}
\right]
+
i\lambda_{2}
\textcolor{purple}{\delta_{Z_{2}}}
\left[
\chi_{ab}
\chi_{cd}
+
\chi_{ad}
\chi_{cb}
\right]
\\
&
+
i\lambda_{3}
\textcolor{purple}{\delta_{Z_{3}}}
\left[
    (S^{\mu\nu})_{ab}
    (S_{\mu\nu})_{cd}
    +
    (S^{\mu\nu})_{ad}
    (S_{\mu\nu})_{cb}
\right]
\\
&
+
i\lambda_{4}
\textcolor{purple}{\delta_{Z_{4}}}
\left[
    (M^{\mu\nu})_{ab}
    (M_{\mu\nu})_{cd}
    +
    (M^{\mu\nu})_{ad}
    (M_{\mu\nu})_{cb}
\right],
\end{split}
\end{equation}
where the coefficients $\{ c_{1}^{1},c_{\chi}^{1},c_{S}^{1},c_{M}^{1},c_{1}^{S},c_{\chi}^{S},c_{S}^{S},c_{M}^{S}\}$ are given by
\begin{equation}
\label{coefficients}
\begin{aligned}
c_{1}^{1}
=&
-11 \lambda_{1}^2-2 \lambda_{1} (8 \lambda_{4}+12 \lambda_{3}+\lambda_{2} )-48 \left(\lambda_{4}^2+2 \lambda_{4} \lambda_{3}+2 \lambda_{3}^2\right)+8 \lambda_{3} \lambda_{2} -3 \lambda_{2} ^2,
\\
c_{\chi}^{1}
=&
-8 \lambda_{2}  (\lambda_{1}+2 \lambda_{4}-2 \lambda_{3})-48 (\lambda_{4}-\lambda_{3})^2-8 \lambda_{2} ^2,
\\
c_{S}^{1}
=&
-8 \lambda_{3} (\lambda_{1}-4 \lambda_{4}+2 \lambda_{3}),
\\
c_{M}^{1}
=&
24 \lambda_{3}^2-8 \lambda_{4} (\lambda_{1}+\lambda_{4}+\lambda_{2} ),
\\
c_{1}^{S}
=&
-16 \left(6 \lambda_{1}^2+\lambda_{1} (18 \lambda_{4}+30 \lambda_{3}+\lambda_{2} )-2 (\lambda_{4}+\lambda_{3}) (4 \lambda_{4}+8 \lambda_{3}-3 \lambda_{2} )\right),
\\
c_{\chi}^{S}
=&
16 \left(\lambda_{2}  (\lambda_{1}+18 \lambda_{4}-30 \lambda_{3})+2 (\lambda_{4}-\lambda_{3}) (3 \lambda_{1}-4 \lambda_{4}+8 \lambda_{3})+6 \lambda_{2} ^2\right),
\\
c_{S}^{S}
=&
-3 \lambda_{1}^2+32 \lambda_{1} \lambda_{4}-2 \lambda_{1} (4 \lambda_{3}+\lambda_{2} )-16 \left(8 \lambda_{4}^2+23 \lambda_{3}^2\right)+8 \lambda_{2}  (4 \lambda_{4}+\lambda_{3})-3 \lambda_{2} ^2,
\\
c_{M}^{S}
=&
48 \lambda_{3} (\lambda_{1}-8 \lambda_{4}+\lambda_{2} ).
\end{aligned}
\end{equation}
Hence, to cancel the divergences in \eqref{VertexDivergence}, the counterterms $\textcolor{purple}{\delta_{Z_{1}}},\textcolor{purple}{\delta_{Z_{2}}},\textcolor{purple}{\delta_{Z_{3}}},\textcolor{purple}{\delta_{Z_{4}}}$, in the MS scheme, must be chosen as
\begin{equation}
\label{VertexCounterterms}
\begin{aligned}
    \textcolor{purple}{\delta_{Z_{1}}}
    =&
    \frac{1}{\lambda_{1}}
    \frac{1}{(4\pi)^{2}
    (\alpha^{2}-\beta^{2})^{2}
    \epsilon}
\left(
\alpha^{2}
c_{1}^{1}
+
\frac{\beta^{2}}{12}
c_{1}^{S}
\right),
&
    \textcolor{purple}{\delta_{Z_{2}}}
    =&
    \frac{1}{\lambda_{2}}
    \frac{1}{(4\pi)^{2}
    (\alpha^{2}-\beta^{2})^{2}
    \epsilon}
\left(
\alpha^{2}
c_{\chi}^{1}
+
\frac{\beta^{2}}{12}
c_{\chi}^{S}
\right),
\\
    \textcolor{purple}{\delta_{Z_{3}}}
    =&
    \frac{1}{\lambda_{3}}
    \frac{1}{(4\pi)^{2}
    (\alpha^{2}-\beta^{2})^{2}
    \epsilon}
\left(
\alpha^{2}
c_{S}^{1}
+
\frac{\beta^{2}}{12}
c_{S}^{S}
\right),
&
    \textcolor{purple}{\delta_{Z_{4}}}
    =&
    \frac{1}{\lambda_{4}}
    \frac{1}{(4\pi)^{2}
    (\alpha^{2}-\beta^{2})^{2}
    \epsilon}
\left(
\alpha^{2}
c_{M}^{1}
+
\frac{\beta^{2}}{12}
c_{M}^{S}
\right).
\end{aligned}
\end{equation}

These results show that the inclusion of all the dimension-4 parity-invariant self-interactions is sufficient for renormalizing this model for any values of $\alpha,\beta$; however, in some cases, not all of them are necessary.
From the expressions \eqref{VertexDivergence} and \eqref{coefficients}, we can find particular theories, defined from the values of $(\alpha,\beta)$ and subsets of the parameter space $\bm{\lambda}=(\lambda_{1},\lambda_{2},\lambda_{3},\lambda_{4})$, that are renormalizable as well. They are listed next.
\begin{itemize}[label=\textbullet]
    \item Theory i) ($\alpha=1,\beta=0$).
In this case, there are four renormalizable theories,
parameterized by
    \begin{align}
    \label{i)subtheory1}
        \bm{\lambda}
        =&
        \left( \lambda_{1},\lambda_{2}=0,\lambda_{3}=0,\lambda_{4}=0 \right),
        \\
    \label{i)subtheory2}
        \bm{\lambda}
        =&
        \left( \lambda_{1},\lambda_{2},\lambda_{3}=0,\lambda_{4}=0 \right),
        \\
    \label{i)subtheory3}
        \bm{\lambda}
        =&
        \left( \lambda_{1},\lambda_{2},\lambda_{3}=0,\lambda_{4} \right),
        \\
    \label{i)subtheory4}
        \bm{\lambda}
        =&
        \left( \lambda_{1},\lambda_{2},\lambda_{3},\lambda_{4} \right).
    \end{align}
    \item Theory ii) ($\alpha=0,\beta=1$).
    In this case, all the self-interactions are necessary; there is one single renormalizable theory, parameterized by
    \begin{align}
    \label{ii)subtheory1}
        \bm{\lambda}
        =&
        \left( \lambda_{1},\lambda_{2},\lambda_{3},\lambda_{4} \right).
    \end{align}
\end{itemize}

\subsection{Beta functions}

From the relations \eqref{lambdas} and \eqref{scale} we have that the bare $\lambda_{X,0}$ and physical $\lambda_{X}$ coupling parameters are related by
\begin{align}
    \lambda_{1,0}=&Z^{-2}Z_{1}\lambda_{1}\mu^{2\epsilon},
    &
    \lambda_{2,0}=&Z^{-2}Z_{2}\lambda_{2}\mu^{2\epsilon},
    &
    \lambda_{3,0}=&Z^{-2}Z_{3}\lambda_{3}\mu^{2\epsilon},
    &
    \lambda_{4,0}=&Z^{-2}Z_{4}\lambda_{4}\mu^{2\epsilon}.
\end{align}
Using these expressions along with the $\textcolor{purple}{\delta_{X}}$ definitions in \eqref{deltas} and the results for $\textcolor{purple}{\delta_{Z}}$ in \eqref{deltaZ=0} and $\textcolor{purple}{\delta_{\lambda_{X}}}$ in \eqref{VertexCounterterms}, we can calculate how each coupling parameter $\lambda_{X}$ runs with energy by computing its corresponding beta function $\beta_{\lambda_{X}}=\mu d\lambda_{X}/d\mu$. The results are
\begin{align}
\label{beta functions}
    \beta_{\lambda_{1}}
    =&
-\frac{1}{24 \pi ^2 \left(\alpha ^2-\beta ^2\right)^2}
\left\{
3 \alpha ^2 
\left[
11 \lambda_{1}^2+2 \lambda_{1} (\lambda_{2}+8 \lambda_{4}+12 \lambda_{3})+3 \lambda_{2}^2-8 \lambda_{2} \lambda_{3}
\right.
\right.
\\
&
\nonumber
\left.
+
48 \left(\lambda_{4}^2+2 \lambda_{4} \lambda_{3}+2 \lambda_{3}^2\right)\right]
+
4 \beta ^2 
\left[
6 \lambda_{1}^2+\lambda_{1} (\lambda_{2}+18 \lambda_{4}+30 \lambda_{3})
\right.
\\
&
\nonumber
\left.
\left.
-
2 (\lambda_{4}+\lambda_{3}) (-3 \lambda_{2}+4 \lambda_{4}+8 \lambda_{3})
\right]
\right\},
\\
\beta_{\lambda_{2}}
=&
-\frac{1}{6 \pi ^2 \left(\alpha ^2-\beta ^2\right)^2}
\left\{
6 \alpha ^2 
\left[
\lambda_{1} \lambda_{2}+\lambda_{2}^2+2 \lambda_{2} (\lambda_{4}-\lambda_{3})+6 (\lambda_{4}-\lambda_{3})^2
\right]
\right.
\\
&
\nonumber
-
\beta ^2 
\left[
2 \left(3 \lambda_{2}^2+9 \lambda_{2} \lambda_{4}-15 \lambda_{2} \lambda_{3}-4 \lambda_{4}^2+12 \lambda_{4} \lambda_{3}-8 \lambda_{3}^2\right)
\right.
\\
&
\nonumber
\left.
\left.
+
\lambda_{1} (\lambda_{2}+6 \lambda_{4}-6 \lambda_{3})
\right]
\right\},
\\
\beta_{\lambda_{3}}
=&
-
\frac{1}{96 \pi ^2 
\left(\alpha ^2-\beta ^2\right)^2}
\left\{
96 \alpha ^2 
\left[
\lambda_{3} (\lambda_{1}-4 \lambda_{4}+2 \lambda_{3})
\right]
+
\beta^{2}
\left[
3 \lambda_{1}^2+3 \lambda_{2}^2
\right.
\right.
\\
&
\nonumber
\left.
\left.
+2 \lambda_{1} (\lambda_{2}+4 (\lambda_{3}-4 \lambda_{4}))-8 \lambda_{2} (4 \lambda_{4}+\lambda_{3})+16 \left(8 \lambda_{4}^2+23 \lambda_{3}^2\right)
\right]
\right\},
\\
\label{beta function 4}
\beta_{\lambda_{4}}
=&
-
\frac{1}{2 \pi ^2 \left(\alpha ^2-\beta ^2\right)^2}
\left\{
2 \alpha ^2 
\left[
3 \lambda_{3}^2-\lambda_{4} (\lambda_{1}+\lambda_{2}+\lambda_{4})
\right]
+
\beta ^2 
\left[
\lambda_{3} (\lambda_{1}+\lambda_{2}-8 \lambda_{4})
\right]
\right\}.
\end{align}
These expressions for the beta functions exactly match those reported in \cite{Rivero-Acosta2020} for self-interactions when setting $m=0$, for $\alpha=1$ and $\beta=0$.

The beta functions \eqref{beta functions}-\eqref{beta function 4} depend on the four parameters $ \bm{\lambda}=(\lambda_{1},\lambda_{2},\lambda_{3},\lambda_{4})$, and so there is room for a large set of fixed points. They are listed next.
\begin{itemize}[label=\textbullet]
    \item Theory i) ($\alpha=1,\beta=0$).
    \begin{itemize}[label=$\circ$]
        \item $\beta_{\lambda_{1}}$. No real fixed points.
        \item $\beta_{\lambda_{2}}$. Defined by
        \begin{equation}
        \lambda_{1}
        =
        -\lambda_{2}
        +
        2 (\lambda_{3}-\lambda_{4})
        -
        \frac{6 (\lambda_{3}-\lambda_{4})^2}{\lambda_{2}}.
        \end{equation}
        \item $\beta_{\lambda_{3}}$. Defined by
        \begin{align}
            \lambda_{3}=&0,
            &
            \text{or}&
            &
            \lambda_{1}
            =&
            4\lambda_{4}-2 \lambda_{3}.
        \end{align}
        \item $\beta_{\lambda_{4}}$. Defined by
        \begin{equation}
            \lambda_{1}
            =
            -(\lambda_{2}+\lambda_{4})+\frac{3 \lambda_{3}^2}{\lambda_{4}}.
        \end{equation}
    \end{itemize}
    There are six independent non-trivial common solutions for $\beta_{\lambda_{2}}$, $\beta_{\lambda_{3}}$, and $\beta_{\lambda_{4}}$, parameterized as
    \begin{align}
    \label{fixedpoints1}
    \bm{\lambda}^{(1)}=&
        \left(\lambda_{1},\lambda_{2}=0,\lambda_{3}=0,\lambda_{4}=0\right),
        \\
    \label{fixedpoints2}
        \bm{\lambda}^{(2)}=&
        \left(\lambda_{1},\lambda_{2}=-\lambda_{1},\lambda_{3}=0,\lambda_{4}=0\right),
        \\
    \label{fixedpoints3}
        \bm{\lambda}^{(3)}=&
        \left(\lambda_{1},\lambda_{2}=-\frac{6}{5}\lambda_{1},\lambda_{3}=0,\lambda_{4}=\frac{1}{5}\lambda_{1}\right),
        \\
        \bm{\lambda}^{(4)}=&
        \left(\lambda_{1},\lambda_{2}=-\frac{8}{7}\lambda_{1},\lambda_{3}=-\frac{1}{14}\lambda_{1},\lambda_{4}=\frac{3}{14}\lambda_{1}\right),
        \\
        \bm{\lambda}^{(5)}=&
        \left(\lambda_{1},\lambda_{2}=-\frac{2}{3}\lambda_{1},\lambda_{3}=-\frac{1}{6}\lambda_{1},\lambda_{4}=\frac{1}{6}\lambda_{1}\right),
        \\
        \bm{\lambda}^{(6)}=&
        \left(\lambda_{1},\lambda_{2}=-\frac{6}{5}\lambda_{1},\lambda_{3}=\frac{1}{10}\lambda_{1},\lambda_{4}=\frac{3}{10}\lambda_{1}\right).
    \end{align}
    In particular, notice that $\bm{\lambda}^{(1)}$, $\bm{\lambda}^{(2)}$ and $\bm{\lambda}^{(3)}$ belong to the theories defined by \eqref{i)subtheory1}, \eqref{i)subtheory2} and \eqref{i)subtheory3}, respectively; while $\bm{\lambda}^{(4)}$, $\bm{\lambda}^{(5)}$ and $\bm{\lambda}^{(6)}$ belong to \eqref{i)subtheory4}.
    \item Theory ii) ($\alpha=0,\beta=1$).
    \begin{itemize}[label=$\circ$]
        \item $\beta_{\lambda_{1}}$. Defined by
        \begin{equation}
        \lambda_{2}
        =
            \frac{-6 \lambda_{1}^2-6 \lambda_{1} (5 \lambda_{3}+3 \lambda_{4})+8 (\lambda_{3}+\lambda_{4}) (2 \lambda_{3}+\lambda_{4})}{\lambda_{1}+6 (\lambda_{3}+\lambda_{4})}.
        \end{equation}
        \item $\beta_{\lambda_{2}}$. Defined by
        \begin{equation}
        \lambda_{1}
        =
        \frac{-6 \lambda_{2}^2 - 6 \lambda_{2} (-5 \lambda_{3}+3 \lambda_{4})+8 (-\lambda_{3}+\lambda_{4})(-2\lambda_{3}+\lambda_{4})}{\lambda_{2}+6 (-\lambda_{3}+ \lambda_{4})}.
        \end{equation}
        \item $\beta_{\lambda_{3}}$. Defined by
        \begin{align}
        \label{iibeta3}
        \bm{\lambda}
        =
        \left(
            \lambda_{1},
            \lambda_{2}=\lambda_{1},
            \lambda_{3}=0,
            \lambda_{4}=\frac{1}{4}\lambda_{1}
        \right).
        \end{align}
        \item $\beta_{\lambda_{4}}$. Defined by
        \begin{align}
        \label{iibeta4}
        \lambda_{3}=&0,
        &
        \text{or}&
        &
        \lambda_{1}
        =&
        8\lambda_{4}
        -
        \lambda_{2}.
        \end{align}
    \end{itemize}
    In this case, each beta function has fixed points, however, there are no non-trivial common solutions; notice that a curious solution arises for $\bm{\lambda}=(\lambda_{1},\lambda_{2}=\lambda_{1},\lambda_{3}=0,\lambda_{4}=\lambda_{1}/4)$, for which $\beta_{\lambda_{3}}$ and $\beta_{\lambda_{4}}$ vanish while $\beta_{\lambda_{1}}$ and $\beta_{\lambda_{2}}$ turn out to be the negative of each other: $\beta_{\lambda_{1}}=-\beta_{\lambda_{2}}=-25\lambda_{1}^{2}/12\pi^{2}$.
\end{itemize}

\section{Summary and conclusions}
\label{section5}

Fields belonging to the $(1,0)\oplus(0,1)$ representation of the RLG have been studied since decades ago for both purposes: phenomenological and pure theoretical explorations, aiming to survey new physics. The free EOMs for these fields, previously studied in the literature, are three: i) a Klein-Gordon-like one, ii) the Joos-Weinberg, and iii) the Shay-Good/Hammer-McDonald-Pursey. Their corresponding QFTs for the massive $m\neq 0$ cases, however, present some issues: while the propagator of ii) has an awkward unphysical pole at $p^{2}=-m^{2}$ yielding an unconventional quantum theory, that of iii) is divergent in the UV region ($p^{2}\gg m^{2}$), resulting in a non-renormalizable theory. For $m=0$, in contrast, the corresponding propagators have none of these issues, opening the door for conventional and renormalizable QFTs. The mass-dimension of these fields is one, allowing the definition of several dimension-4 interactions with the SM fields, as well as dimension-4 self-interactions. Thus, there exists a variety of plausible renormalizable theories, among which the most straightforward and economical are those involving only self-interactions. At the same time, self-interactions may be required for renormalizing interactions with SM fields. This led us to study the renormalization of self-interacting massless theories in this representation.

In order to study the renormalization of the entire class of parity-invariant self-interacting massless theories in the $(1,0)\oplus(0,1)$ representation, we constructed the most general model compatible with these attributes. It consisted of a free Lagrangian that reproduces the massless limits of theories i), ii), and iii) as particular cases; along with an interacting Lagrangian containing all the independent dimension-4 parity-invariant self-interactions available in this representation. These interactions, however, explicitly break the gauge symmetry arising in the free massless case of theory iii), and hence we restricted ourselves to retrieving theories i) and ii). This general model has two superficially divergent diagrams: the self-energy and the vertex. Both were successfully renormalized in the MS scheme with appropriate choices for the counterterms. For i), we found four independent renormalizable theories; while for ii), we found just one, turning out that all of the independent self-interactions are necessary. The beta functions were calculated, and their fixed points identified. In particular, for i) we found that $\beta_{\lambda_{1}}$ has no fixed points at all, while there are six independent non-trivial common fixed points for $\beta_{\lambda_{2}},\beta_{\lambda_{3}},\beta_{\lambda_{4}}$; for ii), although each beta function has fixed points, there are no non-trivial common solutions for the four of them; interestingly, a curious solution emerged at which $\beta_{\lambda_{3}}$ and $\beta_{\lambda_{4}}$ vanish, while $\beta_{\lambda_{1}}$ and $\beta_{\lambda_{2}}$ are the negative of each other.

In conclusion, we found that the entire class of parity-invariant self-interacting massless theories in the $(1,0)\oplus(0,1)$ representation of the RLG, with $|\alpha| \neq |\beta|$, is renormalizable at the one-loop level, and the general model we worked with constitutes a rich theoretical laboratory for renormalization group analysis. This class contains the self-interacting massless Klein-Gordon-like theory, as well as the self-interacting Joos-Weinberg one, which have been previously studied in the literature, as particular cases, but is not limited to them. For the massless case with $|\alpha| = |\beta|$, it is necessary to add gauge-invariant interactions, which usually require the inclusion of additional fields. We will address this topic in a coming article. 






\acknowledgments

A.C. R.-P. acknowledges SECIHTI national scolarships.  C.A. V.-A. is supported by the SECIHTI IxM project 749 and SNII 58928.




\bibliographystyle{JHEP}
\bibliography{biblio.bib}



\end{document}